\documentclass[manuscript]{acmart} 

\usepackage{hyperref}

\AtBeginDocument{%
  }

\copyrightyear{2026}
\acmYear{2026}
\acmMonth{4}
\acmDOI{}
\acmISBN{}

\acmConference[]{CHI '26 Workshops: Herding CATs}{April 13--17}{Barcelona, Spain}

\begin{document}

\title{DrawTalking It Out: Creativity-Support Research as Creative Process Itself}

\author{Karl Toby Rosenberg}
\orcid{0000-0001-7744-5188}
\affiliation{%
  \institution{New York University}
  \country{New York, New York, USA}
}
\email{ktr254@nyu.edu}

\renewcommand{\shortauthors}{Rosenberg et al.}

\begin{abstract}
    I think the creative process in conducting open-ended creativity-adjacent research is itself part of the creative process (including ideation, pivots, and tangents excluded from the final work) that deserves greater emphasis. how might we bridge multiple open-ended projects and serve broader non-technical audiences? We may choose to emphasize interaction techniques rather than singular tools in isolation. We may grow a community based on our shareable experiences in creativity interactions, to draw more complete pictures of how we creatively solve creativity problems (or satisfy curiosities). This may lead to new directions. I briefly elaborate on this using the example of "DrawTalking" a drawing+talking interactions work. DrawTalking itself resulted from a winding creative process exploring spontaneous interactive world-building and storytelling when drawing and talking.
\end{abstract}




\begin{teaserfigure}
\centering
  \includegraphics[width=\textwidth]{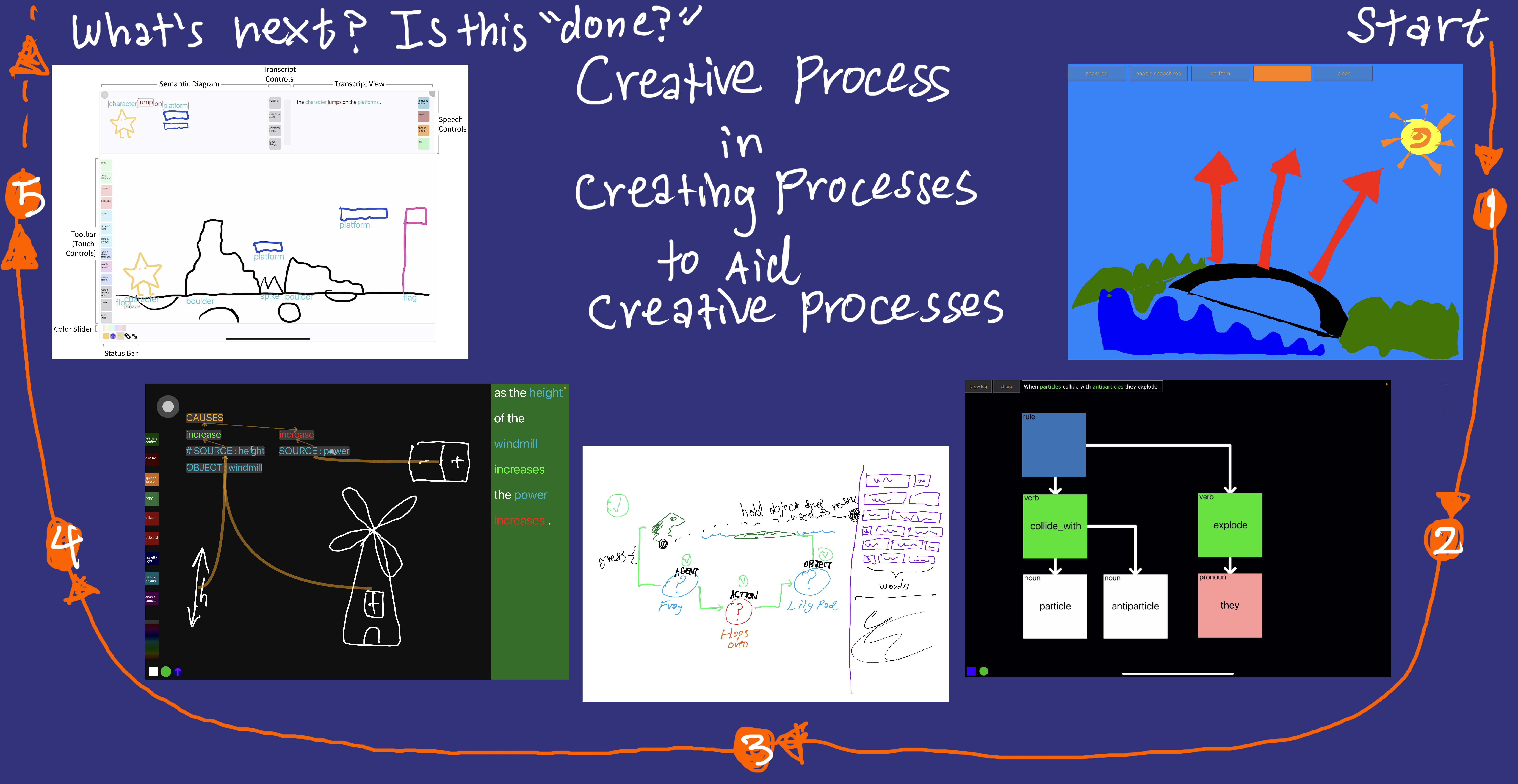}
  \caption{\emph{Creativity-Support Interaction Research as Its Own Creative Process Worth Studying - Selected pieces of making DrawTalking, a drawing+talking interface that shows language elements used in commands the user constructs by talking, to control interactive hand-drawn objects on a canvas}; right-to-left: Start->1): the hobbyist learning to make a sketching application; 2) an unused version of a diagram showing parts-of-speech, self-rejected due to looking too computer science-minded and hard to read; 3) a hand-drawn sketch visualizing a new version of the same diagram that links objects in the canvas to words; 4) an attempted version of the interface implementing v3, still with hard-to-read parts-of-speech that look too close to a computer science-esque tree, and self-rejected due to the diagram getting in the way of the scene; 5) the perhaps-finished revamped interface that isolates the language visuals from the drawn objects more-so, and yet is this really finished, or is it just what was done at the deadline? Who knows if there might have been more iterations or future blends of the different ideas?}
  \Description{Creativity-Support Interaction Research as Its Own Creative Process Worth Studying - Selected pieces of making DrawTalking, a drawing+talking interface that shows language elements used in commands the user constructs by talking, to control interactive hand-drawn objects on a canvas; right-to-left: Start->1): the hobbyist learning to make a sketching application; 2) an unused version of a diagram showing parts-of-speech, self-rejected due to looking too computer science-minded and hard to read; 3) a hand-drawn sketch visualizing a new version of the same diagram that links objects in the canvas to words; 4) an attempted version of the interface implementing v3, still with hard-to-read parts-of-speech that look too close to a computer science-esque tree, and self-rejected due to the diagram getting in the way of the scene; 5) the perhaps-finished revamped interface that isolates the language visuals from the drawn objects more-so, and yet is this really finished, or is it just what was done at the deadline? Who knows if there might have been more iterations or future blends of the different ideas?}
  \label{fig:teaser}
\end{teaserfigure}

\maketitle

\section{Introduction: We Should Treat More of the Creative Process in Open-ended Creativity-Adjacent Research More Directly as Part of the Research}
Creativity-adjacent Human-Computer Interaction research is incredibly broad, but it generally is concerned with the design of tools, exploration of interactions, or analysis of creative processes. However, it can be challenging to figure-out how to define a clear rigorous research path when our goal is somehow loosely-defined and exploratory, towards supporting creativity without a particular audience, quantifiable outcome, comparisons, or measure of success in terms of product output. Oftentimes, the goal is also to support an open-ended form of creativity that lacks a "win" condition. The research itself might be an open-ended exploration personally-motivating to the researcher due to their experience or so new that there isn't really a guaranteed path forwards. A lack of scaffolding on how to approach this research in a way that is scientifically valid can be especially challenging, time-consuming, and demoralizing when met with skepticism. However, open-endedness should be okay since it can help us develop truly new ideas.

Indeed, Takeo Igarashi's UIST '25 visions talk "Easy and Fast"\cite{EasyAndFast} suggests (to paraphrase) that we often may be concerned too much with following a formulaic approach (formative studies to create a motivation that maybe was there from personal experience already, but which "needed" a more "researchy" angle to pass; or feeling a need to measure success by speed/efficiency improvements as opposed to subjective elements like fun), and that we may wish to explore creativity interfaces whose goals are more open-ended and aimed at supporting fun or new experiences. I agree with Professor Igarashi's points in this talk, but to expand, I think we also need to consider our creative process in designing the research tool/interaction as a more significant contribution to the core of the research. When we look at "final" papers and portfolio pieces, all we see is a compressed surface-level version of a given project. This doesn't lend itself well to open-projects, I think, because the result omits the depth and nuances of the full exploration process in figuring-out and iterating on the project. In other words, it likely leaves-out valuable information about the creative process of the research: e.g. the drastic changes, failures, pivots, conversations and online chats with other people, etc. that didn't smoothly fit the narrative of the final project. I believe that in focusing on the research product, we're missing-out. Additionally, by focusing on tool-building and narrow criteria for success, versus open-ended interactions, we're also missing-out on opportunities for sharing research, through the process of creating all-in-one tools (on repeat). 

If our goal collectively as creativity HCI researchers is to build a more collaborative and communicative community, we may increasingly emphasize interaction techniques
rather than singular tools in isolation. On isolated tools: Li et al. argues that truly empowering creative users with the ability to make independent choices means supporting flexible decision-making support in a given tool, as well as inter-operation with external tools \cite{Li-BeyondArtifact-Power-Lens-for-CST}. This also suggests we may want to think in terms of generalizable and complementary interactions, and move away from all-in-one systems (which isn't to say we are only focusing on tools e.g. \cite{Wendy-Michel_InteractionSubstrates, klokmose_webstrates_2015}.) This might also clarify that the point of a lot of open-ended projects is capturing a particular feel/experience independent of high technical and visual fidelity (something not everyone can afford to achieve to highlight their interaction concepts). We may then grow a community based on more easily shareable experiences in creativity interactions, to draw more complete pictures of how we design interfaces (or satisfy curiosities). This may lead to new directions due to less time spent on creating "polished" tools, and more on supporting each other's work. (Perhaps we can reduce emphasis on siloed "apps" \cite{michel_beaudouin-lafon_world_without_apps_2019}). Also, I would advocate for stronger guidelines for conducting open-ended research that clarify to audiences outside our circles what constitutes "good" research. From experience, a lack of clarity in how open-ended design research is accepted in HCI can make it painful to defend against dismissive, perhaps more quantitative-only-minded researchers.

For a concrete example, I'll briefly discuss these points through the lens of of my "DrawTalking:" a drawing+talking interactions work that itself resulted from a winding creative process. It was meant to explore ways we could support spontaneous creative processes for world-building when drawing and talking \cite{DrawTalking}. As the first author, I found DrawTalking to be a rewarding project, but there were unseen revisions and challenges that in retrospect, I felt could've formed a more complete and compelling story. I'll also share some research "brick walls" I hit in my attempts to publish the research, created in part by the open-endedness and lack of one-size-fits-all approaches. I'd like to see it more widely-encouraged to share the complete story of such projects, and for the community to co-create the scaffolding and academic support systems to do so more so in the future, perhaps to make it possible to do open-ended creativity research more easily without fearing as many dead-ends or feeling a need to apply research methods even when they're unnatural. We could have more support-systems in-place to promote and encourage exploratory work.

\section{The Unseen and Informal Making of (and Challenges of Publishing) DrawTalking: Or Should It Be Okay to Be More Informal?}
Here I offer some anecdotes on the research process for DrawTalking including some of the challenges when trying to publish it, coupled with some rhetorical questions I'd like to propose for discussion.

\textbf{Is personal motivation enough to justify creative tools research?} The final paper follows a rather standard process going from formative studies to ground the research in real-world drawing+talking practices at the whiteboard, to design goals, to system design and implementation. Much of the full picture was omitted because it seemed less rigorous or aligned with a smooth version of the story behind the work. Namely, although the formative steps described in the paper were important for developing the research ideas, much of the main system design had already been storyboarded and built-up in parallel with this exploration, not completely in sequence. I'm no longer convinced this is wrong, but it did seem messy. The paper omits the extremely important conversations between the coauthors and myself, as well as personal motivation for working on the project. Specifically, Professor Xia, Perlin, and I simply were brainstorming what-ifs. i.e. "what if we added speech controls to Professor Perlin's existing Chalktalk project \cite{chalktalk_sketch_commands_2018}?" Simultaneously, I personally was interested in learning more graphics and game engine programming, and was long-hoping to take inspiration from the 2.5D level-building, sculpting, and logic-based games in the LittleBigPlanet series \cite{game:LittleBigPlanet, game:LittleBigPlanet2} to make my own interactive world-building tool. The first application I made for the iPad was just a test of some APIs and whether I could render things to the screen for fun. The development process for what became DrawTalking started with these personal tests (see the teaser for some steps \ref{fig:teaser}). None of these steps seems like "professional" research but they were important personal experiences that motivated the design of the system. The context isn't required, and yet I now think it would've been nice to include this piece of my personal creative journey. \textbf{Might we want to be more accepting of informal formative steps in our research's creative process? Where would this fit-in?}

\textbf{DrawTalking actually had two very-different-looking designs, but only the final was described.} (See the teaser for some steps \ref{fig:teaser}). Paper space constraints aside, might we be losing a lot of useful information when discarding major design changes and pivots? The teaser shows how in particular the semantics diagram (see the DrawTalking paper), which conveys the state of the system's natural language understanding to the user, went through several design concepts, starting from a tree-like computer science diagram, towards a more flattened sentence-like visualizations. I felt the final design would be more readable to non-CS people without technical backgrounds, and easier to read for everyone if separated from the canvas instead of being mixed with visual content in the canvas. This evolution was omitted from the paper, and yet I now think it might've been a useful takeaway to share: to show why certain designs might not work for certain audiences, even if they are quote/unquote failures. Another example: "DrawTalking" was originally called "DrawTalk" for an initial unsuccessful submission, but I felt "DrawTalking" better conveyed the focus on creative process and experience versus output and production quality. The initial reviewers were contested the idea that "anyone could tell a story with this" (to paraphrase), which gave me the sense that they fixated too much on how not-pretty/polished the tool was, rather than focusing on the general interactions. In retrospect, having to create an entire tool like this might've taken longer than the research, but there was a constant pressure to make things look good enough to appease reviewers enough to get them to focus on the interactions concepts that mattered. In short: \textbf{Is there a way to de-emphasize tool building and emphasize interactions prototyping? Would sharing larger views into the creative process behind these projects be helpful in getting the point across that "it's not about the fidelity?" I suspect DrawTalking is very much not the only HCI prototype that was at one point blocked by an audience fixated on product or output. It would be nice to have a shortcut around this misconception.}

\textbf{Legitimizing open-ended creativity research is sometimes challenging} if quantitative evaluations (e.g. with direct comparisons and statistics) are expected within communities that are unfamiliar with or reluctant to accept more qualitative methods as rigorous science. (E.g. methods described in \cite{ledo_evaluation_2018, olsen_evaluating_2007}). Comparison studies might sometimes make sense to gauge usability or performance, if these are relevant. However, if there's no reasonable baseline or participants are heavily-biased in favor of baseline workflows from having used them a lot, and/or your use cases are open-ended, you'll get stuck or question if the evaluations are fair to your own work. If a project is an ecosystem with features that can't simply be removed without breaking the whole, then ablation studies won't work either \cite{arawjo_kicking_2023}. For DrawTalking, it made more sense to collect in-depth feedback from playtesting-style sessions and conversations instead. I decided to do just that and talk with people (professors, students) 1-1 prior to the feedback recorded in the paper. I showed my project without explaining much, and asked what they felt it was good for and what it was to them. Collecting informal external conversations and perspectives describing the thing I was making helped me frame the story of the work. In retrospect, this is yet another part of the story that was skipped, for the lack of a proper IRB or process for correctly including this feedback. \textbf{I think we should include plenty more of this kind of informal discussion in relation to our creative work, as part of the research process, but I'm unsure of how to so-to-speak standardize this since it's so ad-hoc. This may make it easier to overcome blockers in open-ended research where maybe the only way forward is through searching for spontaneous conversations and feedback.}

\section{Conclusion}
I think that we ought to grow a creative-tools community that removes friction in the way of open-ended research and which could offer support and mechanisms for researchers to develop their ideas outside the standard research processes. We need more opportunities and flexibility for more forms of research-through design and personal creative processes, and maybe this means encouraging us to show more of what happens behind the scenes during a project, including personal motivation and conversations between people, usually left-out due to seeming too insignificant and informal. I hope the points lead to helpful discussions.

\bibliographystyle{ACM-Reference-Format}
\bibliography{main.bib}

@misc{arawjo_kicking_2023,
    title = {Kicking the {Leg} {Out} {From} the {Table}: {On} {Contrived} {Controls} in {HCI} {Systems} {Research}},
    shorttitle = {Kicking the {Leg} {Out} {From} the {Table}},
    url = {https://ianarawjo.medium.com/kicking-the-leg-out-from-the-table-on-contrived-controls-in-hci-systems-research-9f77b28fef39},
    language = {en},
    urldate = {2026-02-12},
    journal = {Medium},
    author = {Arawjo, Ian},
    month = oct,
    year = {2023},
}

@inproceedings{EasyAndFast,
author = {Igarashi, Takeo},
title = {Easy and fast? Rethinking The future of content creation tools},
year = {2025},
isbn = {9798400720369},
publisher = {Association for Computing Machinery},
address = {New York, NY, USA},
url = {https://doi.org/10.1145/3746058.3762828},
doi = {10.1145/3746058.3762828},
booktitle = {Adjunct Proceedings of the 38th Annual ACM Symposium on User Interface Software and Technology},
articleno = {17},
numpages = {2},
location = {
},
series = {UIST Adjunct '25}
}

@inproceedings{DrawTalking,
author = {Rosenberg, Karl Toby and Kazi, Rubaiat Habib and Wei, Li-Yi and Xia, Haijun and Perlin, Ken},
title = {DrawTalking: Building Interactive Worlds by Sketching and Speaking},
year = {2024},
isbn = {9798400706288},
publisher = {Association for Computing Machinery},
address = {New York, NY, USA},
url = {https://doi.org/10.1145/3654777.3676334},
doi = {10.1145/3654777.3676334},
booktitle = {Proceedings of the 37th Annual ACM Symposium on User Interface Software and Technology},
articleno = {76},
numpages = {25},
location = {Pittsburgh, PA, USA},
series = {UIST '24}
}

@misc{michel_beaudouin-lafon_world_without_apps_2019,
    address = {ACM UIST},
    type = {{YouTube} {Video}},
    title = {A {World} {Without} {Apps}},
    url = {https://www.youtube.com/watch?v=ntaudUum06E},
    author = {{Michel Beaudouin-Lafon}},
    year = {2019},
    note = {Recorded at the ACM Symposium on User Interface Software and Technology (UIST) 2019 in New Orleans, LA, USA, October 20-23, 2019.},
}

@inproceedings{Wendy-Michel_InteractionSubstrates,
author = {Mackay, Wendy E. and Beaudouin-Lafon, Michel},
title = {Interaction Substrates: Combining Power and Simplicity in Interactive Systems},
year = {2025},
isbn = {9798400713941},
publisher = {Association for Computing Machinery},
address = {New York, NY, USA},
url = {https://doi.org/10.1145/3706598.3714006},
doi = {10.1145/3706598.3714006},
booktitle = {Proceedings of the 2025 CHI Conference on Human Factors in Computing Systems},
articleno = {687},
numpages = {16},
location = {
},
series = {CHI '25}
}

@inproceedings{Li-BeyondArtifact-Power-Lens-for-CST,
author = {Li, Jingyi and Rawn, Eric and Ritchie, Jacob and Tran O'Leary, Jasper and Follmer, Sean},
title = {Beyond the Artifact: Power as a Lens for Creativity Support Tools},
year = {2023},
isbn = {9798400701320},
publisher = {Association for Computing Machinery},
address = {New York, NY, USA},
url = {https://doi.org/10.1145/3586183.3606831},
doi = {10.1145/3586183.3606831},
booktitle = {Proceedings of the 36th Annual ACM Symposium on User Interface Software and Technology},
articleno = {47},
numpages = {15},
location = {San Francisco, CA, USA},
series = {UIST '23}
}

@inproceedings{klokmose_webstrates_2015,
    address = {New York, NY, USA},
    series = {{UIST} '15},
    title = {Webstrates: {Shareable} {Dynamic} {Media}},
    isbn = {978-1-4503-3779-3},
    shorttitle = {Webstrates},
    url = {https://doi.org/10.1145/2807442.2807446},
    doi = {10.1145/2807442.2807446},
    urldate = {2023-02-26},
    booktitle = {Proceedings of the 28th {Annual} {ACM} {Symposium} on {User} {Interface} {Software} \& {Technology}},
    publisher = {Association for Computing Machinery},
    author = {Klokmose, Clemens N. and Eagan, James R. and Baader, Siemen and Mackay, Wendy and Beaudouin-Lafon, Michel},
    month = nov,
    year = {2015},
    pages = {280--290},
}

@inproceedings{ledo_evaluation_2018,
    address = {New York, NY, USA},
    series = {{CHI} '18},
    title = {Evaluation {Strategies} for {HCI} {Toolkit} {Research}},
    isbn = {978-1-4503-5620-6},
    url = {https://doi.org/10.1145/3173574.3173610},
    doi = {10.1145/3173574.3173610},
    urldate = {2023-02-20},
    booktitle = {Proceedings of the 2018 {CHI} {Conference} on {Human} {Factors} in {Computing} {Systems}},
    publisher = {Association for Computing Machinery},
    author = {Ledo, David and Houben, Steven and Vermeulen, Jo and Marquardt, Nicolai and Oehlberg, Lora and Greenberg, Saul},
    month = apr,
    year = {2018},
    pages = {1--17},
}

@inproceedings{olsen_evaluating_2007,
    address = {New York, NY, USA},
    series = {{UIST} '07},
    title = {Evaluating user interface systems research},
    isbn = {978-1-59593-679-0},
    url = {https://doi.org/10.1145/1294211.1294256},
    doi = {10.1145/1294211.1294256},
    urldate = {2023-02-20},
    booktitle = {Proceedings of the 20th annual {ACM} symposium on {User} interface software and technology},
    publisher = {Association for Computing Machinery},
    author = {Olsen, Dan R.},
    month = oct,
    year = {2007},
    pages = {251--258},
}

@misc{game:LittleBigPlanet,
    key = "Little Big Planet",
    title         = "{L}ittle {B}ig {P}lanet",
    organization  = "Media Molecule",
    location      = "United Kingdom",
    howpublished  = "[Playstation 3]",
    year          = "2008",
    url = {https://www.mediamolecule.com/games/littlebigplanet}
}

@misc{game:LittleBigPlanet2,
    key = "Little Big Planet 2",
    title         = "{L}ittle {B}ig {P}lanet 2",
    organization  = "Media Molecule",
    location      = "United Kingdom",
    howpublished  = "[Playstation 3]",
    year          = "2011",
    url = {https://www.mediamolecule.com/games/littlebigplanet2}
}

@misc{chalktalk_sketch_commands_2018,
    title = {Chalktalk : {A} {Visualization} and {Communication} {Language} -- {As} a {Tool} in the {Domain} of {Computer} {Science} {Education}},
    shorttitle = {Chalktalk},
    url = {http://arxiv.org/abs/1809.07166},
    doi = {10.48550/arXiv.1809.07166},
    urldate = {2024-01-05},
    publisher = {arXiv},
    author = {Perlin, Ken and He, Zhenyi and Rosenberg, Karl},
    month = sep,
    year = {2018},
    note = {arXiv:1809.07166 [cs]},
}

\end{document}